# Analysis and Evaluation of the Link and Content Based Focused Treasure-Crawler

Ali Seyfi [1,2]

[1]Department of Computer Science
School of Engineering and Applied Sciences
The George Washington University
Washington, DC, USA.

[2]Centre of Software Technology and Management (SOFTAM)
School of Computer Science
Faculty of Information Science and Technology
Universiti Kebangsaan Malaysia (UKM)
Kuala Lumpur, Malaysia

seyfi@gwu.edu

**Abstract**— Indexing the Web is becoming a laborious task for search engines as the Web exponentially grows in size and distribution. Presently, the most effective known approach to overcome this problem is the use of focused crawlers. A focused crawler applies a proper algorithm in order to detect the pages on the Web that relate to its topic of interest. For this purpose we proposed a custom method that uses specific HTML elements of a page to predict the topical focus of all the pages that have an unvisited link within the current page. These recognized on-topic pages have to be sorted later based on their relevance to the main topic of the crawler for further actual downloads. In the Treasure-Crawler, we use a hierarchical structure called the T-Graph which is an exemplary guide to assign appropriate priority score to each unvisited link. These URLs will later be downloaded based on this priority. This paper outlines the architectural design and embodies the implementation, test results and performance evaluation of the Treasure-Crawler system. The Treasure-Crawler is evaluated in terms of information retrieval criteria such as recall and precision, both with values close to 0.5. Gaining such outcome asserts the significance of the proposed approach.

—————— ◆ ——————

## 1 INTRODUCTION

To improve the quality of searching and indexing the Web, our proposed focused crawler depends on two main objectives, namely, to predict the topic of an unvisited page, and to prioritize the unvisited URLs within the current page by using a data structure called T-Graph. We elaborated the architecture of the Treasure-Crawler in a preceding paper [1], where a review on this subject field was discussed by naming and briefly describing some significant Web crawlers. Also, the requirements of a focused crawler were elicited and the evaluation criteria were outlined.

Based on the current enormous size of the Web, which has passed twenty billion Web pages [2], the need for the distribution of resources to harvest, download, store and index the Web has become a clear necessity. In this context, focused crawlers are inevitable tools, aimed for detecting pages based on the diverse topics of human knowledge, and indexing the relevant pages into distributed repositories while keeping a merged index for further faster retrievals.

This task of indexing must be capable enough to keep up with the changes of dynamic contents on the Web, since keeping an old version of the Web is obsolete. That is the reason for employment of machine learning or reinforcement learning strategies within the body of some search engines; to diligently watch the Web. Following is a brief introduction to our proposed prediction method and evaluation criteria. Then the implementation and experimental setup details and results are discussed. This paper concludes with the evaluation of the results and the future directions for interested researchers.

### 1.1 Research Methodology

In order to enhance the accuracy of detecting on-topic pages on the Web, in this experiment we propose a custom method that basically relies on the surrounding text of a URL. As the first objective, we utilize the Dewey Decimal Classification system to pre cisely predict and categorize the topic of an unvisited page. For the second objective: the use of a flexible hierarchical structure of exemplary documents will accelerate the crawler to locate the highly populated on-topic regions of the Web. Based on these objectives, we designed and implemented our system as a prototype. From the evaluated results of our experiment, we can assert that the framework, architecture, design, and implementation of our prototype were effective and satisfied the two objectives of this research. The rest of this paper briefly introduces the framework and architecture of the Treasure-Crawler system, followed by the experiment and its results.

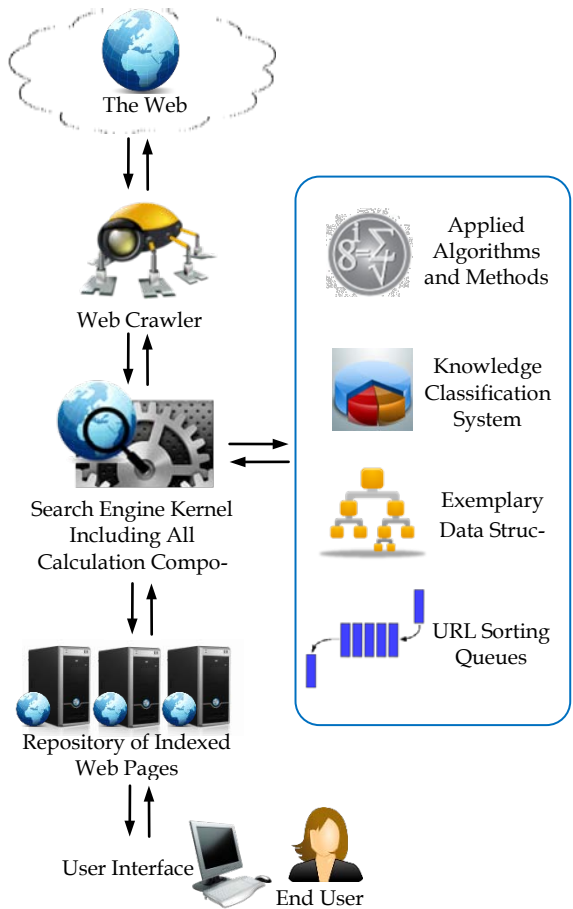

Fig. 1. Conceptual framework of the Treasure-Crawler-based search engine

## 2. FRAMEWORK AND ARCHITECTURE OF THE TREASURE-CRAWLER SYSTEM

Fig. 1 illustrates the framework architecture of a Treasure-Crawler-based search engine and its modules which satisfy all the functional and non-functional requirements. The details of this architecture are given in an under-review paper titled as: "A Focused Crawler Combinatory Link and Content Model Based on T-Graph Principles" [1]. These modules are designed in a way to have the ability of being plugged and played while requiring minimum changes in other modules or the adjacent module interfaces.

The functions of the Treasure-Crawler modules are:

**Web Crawler** is a software agent that receives a URL from the URL's queue (supervised by the kernel) and downloads its corresponding Web page. The page is then parsed and analyzed by the kernel to be ready for the next stages. If the crawler receives any other HTTP response instead of the page content, it tries to store the URL along with its specific message.

**Search Engine Kernel** is the main computational component of the search engine. All the decision makings and necessary calculations are performed in the kernel. It controls all other components while utilizing their data. In focused crawling, the relevancy of the pages and priority assignment to each URL are the main duties of the kernel. In our proposed architecture, the kernel predicts the topical focus of an unvisited link through analyzing its anchor text and the content of the parent page. Thus, the system is set to combine link-based and content-based analysis.

**Applied Algorithms and Methods** are variously utilized to achieve the ultimate goal of the system, which is to effectively index and search the Web. In our system, these custom methods are designed to be simple while efficient. Alternatively, one may use other predefined algorithms such as specific classifiers and text similarity measurements.

**Knowledge Classification System** is used as a reference to classify each Web page or unvisited URL into its appropriate sub-category of human knowledge. In the context of information retrieval and library science there are several classifications systems that are widely used. For our system we chose Dewey Decimal system according to its wide range of knowledge coverage.

**Exemplary Data Structure** is a reference hierarchy based on which the priority of the URLs can be calculated. This reference hierarchy is basically a data structure of some sample pages. Similar to our utilized T-Graph, one can design and use a custom structure for this purpose.

**URL Sorting Queues** are key data structures of any search engine, wherein URLs are lined to be utilized in the intended manner. They are mostly organized as priority queues; however first-in-first-out structure is also widely used.

**Repository** stores an indexed version of the Web. Although this repository cannot keep the live version of the Web, it must be indexed for fast retrieval. Also the repository should stay as up-to-date as possible by use of an intelligent and fast module.

**User Interface** is the only user-side component of a search engine. This interface should be graphically designed in such a way to receive the search query easily and present the search results quickly and clearly.

Given the above framework, the architectural design of the Treasure-Crawler search system is shown in Fig. 2, followed by the descriptions of the processes in a sequential order.

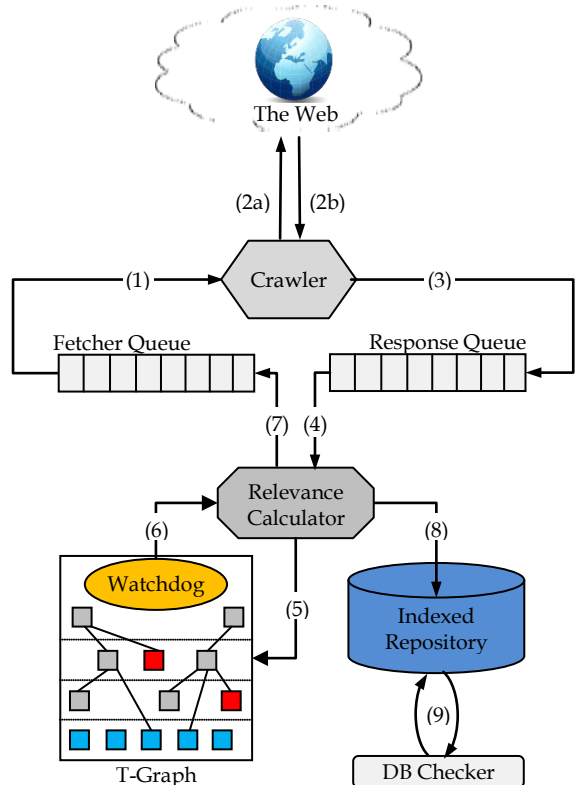

Fig. 2. Architectural design of the Treasure-Crawler Web search system

1. The crawler component dequeues an element from the fetcher queue, which is a priority queue. Initially the seed URLs are inserted into the queue with the highest priority score. Afterwards, the items are dequeued on a highest-priority-first basis.
2. (a) The crawler component locates the Web documents pointed by the URL taken from the fetcher queue and (b) attempts to download the actual HTML data of the page, or otherwise, the server's HTTP response.
3. For each document downloaded, the crawler places the response in the response queue. The response queue contains the documents or HTTP response, in case the page cannot be downloaded due to temporary unavailability or non-freshness of the link.
4. The document is then handed to the relevance calculator which processes the documents and analyzes whether the document belongs to the specialized topic or not.
5. If considered as on-topic, particular elements of the page are then sent to the T-Graph to carry out specific comparisons and calculations. The T-Graph data is used to determine the importance of the unvisited links within the page.
6. T-Graph associates a priority score to each unvisited link. Even the off-topic URLs are assigned a lowest value as the priority. This helps the crawler to harvest unlinked on-topic regions on the Web, which are connected through off-topic pages. There is an observing component called the watchdog on the T-Graph. It periodically updates the T-Graph in order to incorporate its experiences.
7. After completing all analyzes, the relevance calculator inserts the URLs and their assigned priority scores to the fetcher queue. The priority score of the fetcher queue items is cyclically incremented to prevent starvation.
8. HTML data of the analyzed Web page is fully stored in the repository along with all the measurements such as the priority scores given to the links.
9. A database checker component constantly runs specific checkups on the repository and updates its indexes with the ultimate goal of keeping the repository up to date.

## 3. PRACTICAL PREDICTION METHOD – A SCENARIO USING DEWEY DECIMAL CLASSIFICATION (DDC) SYSTEM

Detecting the topical focus of an unvisited page and assigning appropriate priority score to its URL are the main functions of a focused crawler. For the first task, we designed a custom method to detect the main subject(s) of the target page, described as follows:

### 3.1 Detecting the Topical Focus

In order to best classify the main subject of a Web page into known and standard human knowledge categories, we exploit the Dewey decimal system [3, 4], which has 10 super-categories, each of which with 10 categories and again each with 10 sub-categories. These three digits are followed by a decimal point and the rest of the numberings take the subject into more details and specificity. In our experiment, we depended only on the first 3 digits before the decimal point, while the system is implemented in such a way to accommodate any number of digits. By convention, each code of the Dewey system is called a D-number.

Starting from the words of an unvisited link in the current Web page, a (conceptual) 2-D graph is plotted with the complete list of D-Numbers and their length as the dimensions (see Fig. 3). Before any textual comparison, the text is stemmed using the Porter stemming algorithm [5]. This way, the textual data is normalized and the errors are minimized.

Although there are many methods in the Data Mining subject field to eliminate the outliers from a sample data set [6], in our system a simpler method is used to detect the Galaxy, while taking several factors into account. This algorithm takes three characteristics of the words present in the Galaxy. First is the number of words which determines the thickness of the population. Second is the D-Number length which explains the preciseness of the topic. Third is called the anchor impact, where black colored dots have a higher impact since they belong to the words in the anchor text. These words have an effect of about 40% more than the text around the unvisited link [7]. Fig. 3 shows an example of how a graph looks when plotted along with the explanation of the concept as follows:

• All the dots constitute the words between the break points.
• Each light colored (un-bolded black or grayish) dot shows the point corresponding to the word, which is not the anchor text. Each dark colored (bolded black) dot shows the point corresponding to the anchor text of the unvisited link.
• The target is to find the area with the highest population of dots called the Galaxy, which will show the topical focus of the text.

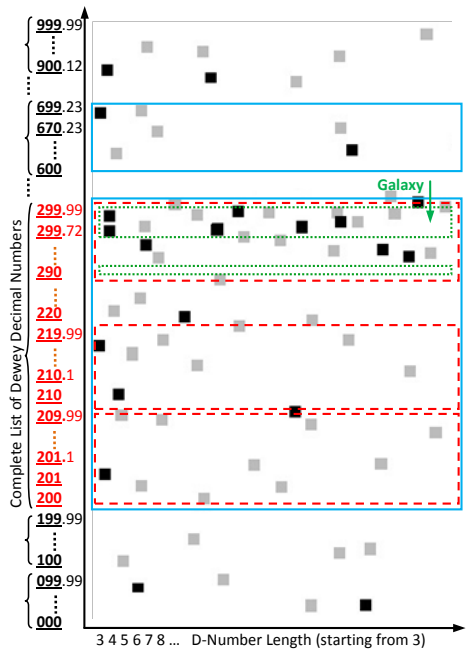

Fig. 3. Example of processing the D-Numbers and retrieving the galaxy

In order to segregate the textual data of the Web page into words/phrases, an HTML document parser is used. It gives phrases if any phrase is directly present in DDC, otherwise it gives the words. It is possible for a word/phrase to have more than one D-Number. The word "clothing" is an example which belongs to several disciplines. It belongs in 155.95 for its psychological meaning; customs associated with clothing belong in 391 as part of the discipline of customs; and clothing in the sense of fashion design belongs in 746.92 as part of the discipline of the arts. Therefore, for each word/phrase, more than one point may be plotted. The plotting is stopped when the paragraph boundary is reached (called break points, which initially could constitute the starting and ending of the paragraph itself). If the unvisited link is a list item, then also the plotting is stopped after plotting all the list items. Then the plotted points are analyzed. [8]

In addition to the above procedure, the T-Graph structure as an exemplary guide carries out the task of priority association by providing a conceptual route for the crawler to follow and find on-topic regions. This phase is elaborated in [1].

### 3.2 Evaluation Criteria

A search engine is designed to retrieve huge amounts of information by surfing or crawling the Web in an automated manner. However, both search engines and traditional information retrieval systems share common parameters in specifying the evaluation criteria. Besides, in a search engine some additional operational parameters are determinant such as the volume and the distribution of the Web, as well as the software architecture and network-related attributes. Considering the essence of our system prototype, we evaluate the Treasure-Crawler based on the criteria that relate to the accuracy of the system. Following is a list of important criteria that should be applied in the evaluation of an ideal focused crawler:

**Harvest ratio:** This is the rate at which relevant pages are detected and high-prioritized, and irrelevant pages are efficiently discarded or low-prioritized. This value should reach a threshold; otherwise an exhaustive crawler may be preferably used.

**Recall:** This is the proportion of relevant retrieved Web pages to all the relevant pages on the Web. To approximate the system recall, we presumed our crawled pages as the whole collection and only took the pages that were topically known within the DMOZ dataset [9].

**Precision:** This is the proportion of relevant retrieved Web pages to all the retrieved Web pages. This metric should also be high.

**Performance:** This is a general term to use but what is important is that the system should have good performance in terms of speed and other IR accuracy metrics.

**Scalability:** In any scale, from a small dataset to the huge Internet, the system should be adjustable and scalable. Our prototype is implemented to scale the Web as well as any provided dataset of documents and URLs, such as the DMOZ, on which it has been run and tested. However it is not evaluated against this criterion because the results of such examination will not be substantial unless actual parallel Web servers and high bandwidth Internet connection are available.

**Modularity:** The system should be implemented by using as many standards as possible, for providing flexible and an easy way for further development on interfaces and mechanisms. In the case of replacement of any components of the system, the other parts should require the least change. Although this criterion is not evaluated in detail, the prototype system of this study is specified, designed and implemented to follow software engineering modularity philosophy whereby one module can be easily replaced with another new one without affecting any other modules of the system. Since the internal working functions of the modules are self-contained, any required changes could be only at the interface between modules if a new or unwanted parameter exchange demands it. As we are following the object oriented approach, the classes and their objects are designed and interconnected in such a way to flexibly accommodate changes and replacements.

**Robustness:** The system has to tolerate corrupted HTML data, unexpected server configurations and behaviors and any other strange issues and errors. Our system detects and records these situations and makes specific decisions or provides error messages for the system administration function.

Our proposed system is implemented as a prototype and tested on a typical machine with a permanent high speed Internet connection. The network, hardware and software conditions of such experimental prototype limited our evaluations to focus on the information retrieval concepts, hence only those criteria that represent the accuracy of the retrieval will be measured and presented. It must be stated that the prototype is designed and implemented in such a way to satisfy those criteria that relate to its software engineering fundaments (e.g. scalability, modularity, and robustness).

## 4. EXPERIMENTAL SETUP

The Treasure-Crawler is designed and implemented based on the modularity and object oriented concepts and techniques; hence all the modules can simply be plugged and played while requiring the minimum change of other modules and at interface level only. The prototype experiment is conducted to test the capability of the crawler to download Web pages related to a predefined topic/ domain, by best using the T-Graph structure and other methods that are defined or used in this research. With this overview, the Treasure-Crawler system was setup as elaborated in Fig. 4, showing the main components as well as the data that travel within the system. The rest of this section is on the description of how the test is carried out.

### 4.1 T-Graph

As described before, the T-Graph has the possibility of being constructed either top-down or bottom-up. According to the requirements of our implementation, the construction of T-Graph starts from a set of target documents which are the most topically relevant Web pages, and from that point onwards, their parents (in terms of Web hyperlinks) are retrieved. Therefore, the utilized approach for T-Graph construction is bottom-up. The target documents in out experiment are:
• http://en.wikipedia.org/wiki/Language
• http://en.wikipedia.org/wiki/Grammar
• http://www.reference.com/browse/language
• http://www.reference.com/browse/grammar

Using one of the Yahoo search tools, the Site Explorer [10], the parents of these four target documents are retrieved to form the nodes of level 1. In the same fashion, the nodes of levels 2 and 3 are retrieved and for each node the constituent components are filled with the necessary data.

In total, the graph has fifty nodes, made out of twenty eight URLs. These nodes constitute the graph in four levels (0-3). It must be stated that several edges are discarded in this implementation of the graph. Table 1 lists the URLs that are used in the T-Graph, as well as number of nodes made out of each.

### 4.2 Topics

The system has been supplied with the DDC codes relevant to the topic of English language and grammar, in which the system specializes. All the textual data is changed to the lower case (although all the comparisons are case insensitive) and stemmed using the Porter stemming algorithm before taking part in any calculation or comparison. This way, the acronyms of each word will be taken as similar to the base word. Table 2 lists the DDC data indicating the topical focus of the system.

### 4.3 Seeds

Seeds are the designated starting points of the system for crawling. These points, in terms of URLs, are initially handed to the crawler to start populating its document collection. These seed URLs as well as the topics of interest constitute the essential parameters of a Web crawler.

In our experiments, two different sets of seeds were supplied to the system. The first set consisted of seventeen on-topic URLs, while the second set contained seven generic URLs, as listed in Table 3.

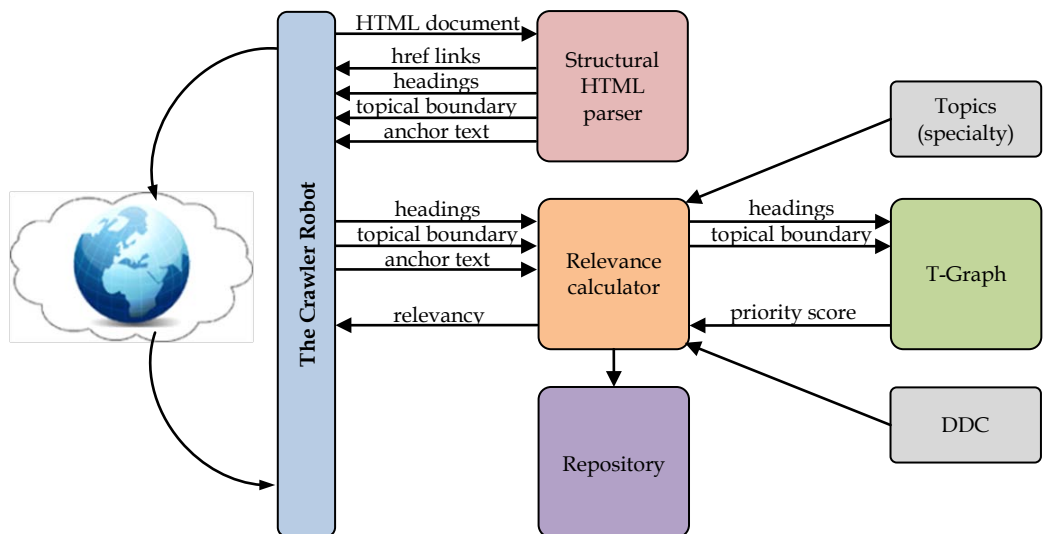

Fig. 4. Crawler robot setup for the experiment

TABLE 1

URLs USED IN OUR ACTUAL T-GRAPH AND THEIR CORRESPONDING NUMBER OF NODES

| No. | URL | Node Count |
|---|---|---|
| 1 | en.wikipedia.org/wiki/Language | 1 |
| 2 | en.wikipedia.org/wiki/Grammar | 1 |
| 3 | www.reference.com/browse/language | 1 |
| 4 | www.reference.com/browse/grammar | 1 |
| 5 | en.wikipedia.org/wiki/Linguistics | 2 |
| 6 | en.wikipedia.org/wiki/Translation | 1 |
| 7 | en.wikipedia.org/wiki/Natural_language | 2 |
| 8 | en.wikipedia.org/wiki/Sentence_(linguistics) | 2 |
| 9 | en.wikipedia.org/wiki/Word | 2 |
| 10 | en.wikipedia.org/wiki/Poetry | 2 |
| 11 | www.reference.com/browse/linguistics | 1 |
| 12 | www.reference.com/browse/speech | 1 |
| 13 | en.wikipedia.org/wiki/Syntax | 2 |
| 14 | en.wikipedia.org/wiki/Drama | 2 |
| 15 | en.wikipedia.org/wiki/Literature | 2 |
| 16 | en.wikipedia.org/wiki/Writing | 2 |
| 17 | en.wikipedia.org/wiki/Etymology | 2 |
| 18 | en.wikipedia.org/wiki/Philosophyoflanguage | 2 |
| 19 | www.reference.com/browse/anthropology | 1 |
| 20 | www.reference.com/browse/mouth | 1 |
| 21 | en.wikipedia.org/wiki/Communication | 4 |
| 22 | en.wikipedia.org/wiki/Sign_language | 4 |
| 23 | en.wikipedia.org/wiki/The_arts | 2 |
| 24 | en.wikipedia.org/wiki/Programming_langua | 3 |
| 25 | en.wikipedia.org/wiki/Anthropology | 2 |
| 26 | en.wikipedia.org/wiki/Genre | 2 |
| 27 | www.reference.com/browse/ethnology | 1 |
| 28 | www.reference.com/browse/dentistry | 1 |

TABLE 2

BASE DDC CODES AND CLASSES

| No. | Code | Class (Initial) |
|---|---|---|
| 1 | 400 | Language |
| 2 | 403 | Dictionaries & encyclopedias |
| 3 | 410 | Linguistics |
| 4 | 412 | Etymology |
| 5 | 413 | Dictionaries |
| 6 | 414 | Phonology & phonetics |
| 7 | 415 | Grammar |
| 8 | 417 | Dialectology & historical linguistics |
| 9 | 418 | Standard usage & applied linguistics |
| 10 | 419 | Sign languages |
| 11 | 420 | English & Old English |
| 12 | 421 | English writing system & phonology |
| 13 | 422 | English etymology |
| 14 | 423 | English dictionaries |
| 15 | 425 | English grammar |
| 16 | 427 | English language variations |
| 17 | 428 | Standard English usage |
| 18 | 429 | Old English (Anglo-Saxon) |
| 19 | 490 | Other Languages |
| 20 | 820 | English & old English literatures |

TABLE 3

ON-TOPIC AND GENERIC SEED URLS

| NO. | URL | Generic(G)/On-Topic(T) |
|---|---|---|
| 1 | http://www.englishclub.com/... | T |
| 2 | http://jc-schools.net/... | T |
| 3 | http://www.brighthub.com/... | T |
| 4 | http://www.englishforum.com/... | T |
| 5 | http://tls.vu.edu.au/...... | T |
| 6 | http://onlineenglishhub.blogspot.com/ | T |
| 7 | http://openlearn.open.ac.uk/...... | T |
| 8 | http://englishhubonline.net/... | T |
| 9 | http://muse.jhu.edu/... | T |
| 10 | http://www.mla.org/... | T |
| 11 | http://robin.hubpages.com/... | T |
| 12 | http://www.scientificpsychic.com/... | T |
| 13 | http://home.pacific.net.au/... | T |
| 14 | http://english.berkeley.edu/... | T |
| 15 | http://englishplus.com/... | T |
| 16 | http://esl.about.com/...... | T |
| 17 | http://www.onlineenglishdegree.com/... | T |
| 1 | http://www.dmoz.org/ | G |
| 2 | http://dir.yahoo.com/ | G |
| 3 | http://vlib.org/ | G |
| 4 | http://www.stpt.com/... | G |
| 5 | http://www.joeant.com/ | G |
| 6 | http://botw.org/ | G |
| 7 | http://www.ansearch.com/... | G |

### 4.4 Experimental Environment

The system is run with different sets of seeds, configurations and conditions in order to be well evaluated. All the runs are carried out on commonly configured machines with permanent Internet ADSL connection. As described before, our experimental prototype system is specialized in the topic of English language and grammar. The April 2011 version of DMOZ directory [9] is employed as the dataset, with the metadata shown in Table 4. It must be stated that the DMOZ dataset is originally an ontology. For our system, the converted version of the DMOZ dumps has been used, which is in MySql database format.

TABLE 4

METADATA OF DMOZ DATASET

| Description | Record Count |
|---|---|
| Total number of URLs | 3,922,814 |
| Total number of topics | 771,986 |
| Number of on-topic URLs | 10,404 |
| Number of related topics | 1,345 |

### 4.5 Running Conditions

While all the parameters and variables of the system has already been calculated and decided, each important one has been alternatively tested under different conditions to come up with an optimum output. Table 5 describes the initial value for each input parameter. The first 3 items of this table have already been described. The 4th item specifies the number of digits of DDC codes (D-numbers) that the system processes to determine the topical focus. Obviously, as the number of processed digits becomes greater, the accuracy of the topic detection increases. The 5th item is the default assigned priority value to an unvisited link when it has no corresponding node in the T-Graph with OSM value of more than the threshold or when the topical focus of the unvisited link is not what the system specializes in. The 6th item is the value that is periodically added to the priority score of each item in the fetcher queue until it is less than 1 to prevent aging of the queue items (also called starvation). This value is added, after a specific number of insertions which is the 7th item of the table.

TABLE 5

INITIAL PARAMETERS DEFAULT VALUES

| No. | Parameter | Value |
|---|---|---|
| 1 | T-Graph Depth | 3.00 |
| 2 | OSM Threshold | 0.05 |
| 3 | Anchor Text Impact Factor | 1.40 |
| 4 | Maximum D-number Length | 3.00 |
| 5 | Unrelated Priority | 0.01 |
| 6 | Fetcher Queue Aging Factor | 0.05 |
| 7 | Item Count to Apply Aging Factor | 100.00 |

As described above, while evaluating the performance of the Treasure-Crawler, we also tested the system against various parameters to observe their affect on the performance. Table 6 lists the conditions under which the system has been run and tested, based on different inputs.

TABLE 6

RUNNING CONDITIONS OF TREASURE-CRAWLER SYSTEM

| On-Topic Seeds | Generic Seeds | Parameter | Value |
|---|---|---|---|
| T1 | G1 | *All defaults* | N/A |
| T2 | G2 | OSM threshold | 0.10 |
| T3 | G3 | Anchor text impact | 0.50 |
| T4 | G4 | Fetcher aging factor | 0.02 |

## 5. EXPERIMENTAL RESULTS OF TREASURE-CRAWLER

In our experiment, the fundamental criterion to be measured for the crawler was the number of retrieved on-topic pages during the runs. Concurrently, we tested the system against diverse input values (as in Table 6) in order to improve our future versions of the Treasure-Crawler. It must be stated that not all the input parameters and their corresponding alternate values are given here in order to prevent the overemphasis on the system parameters. One other fact is that the performance results of the Treasure-Crawler are presented only based on the default values of the input parameters, while we observed that for some parameters, the alternate value yields a better performance. These values will be considered in further implementations of the Treasure-Crawler and presented in follow up papers.

Fig. 5 shows the number of retrieved on-topic pages for runs with generic seed URLs and in each block of 1000 crawled pages. Fig. 6 shows the same result for runs with on-topic seed URLs. Based on these results, it is observed that the number of retrieved pages for runs with on-topic seeds stays more satisfactory. This is because the crawler initially learns to discard the regions on the Web that yield the least number of relevant pag-

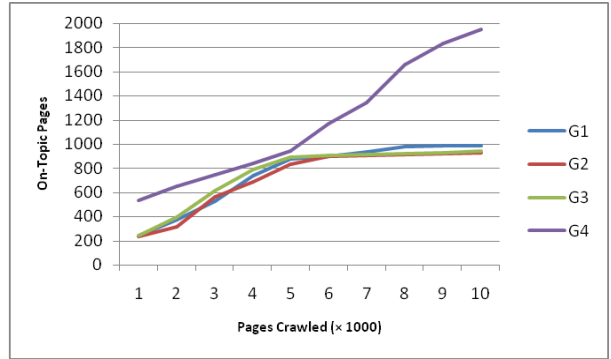
Fig. 7. On-topic retrieved pages (Generic seeds)

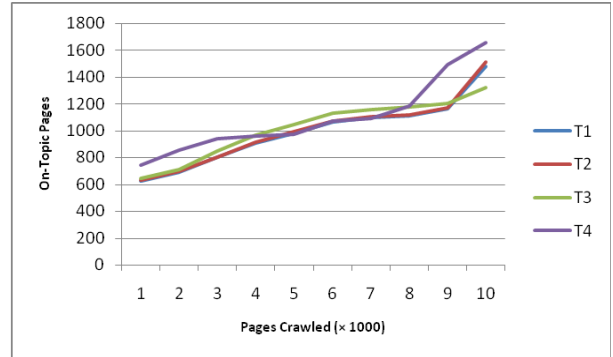
Fig. 8. On-topic retrieved pages (On-topic seeds)

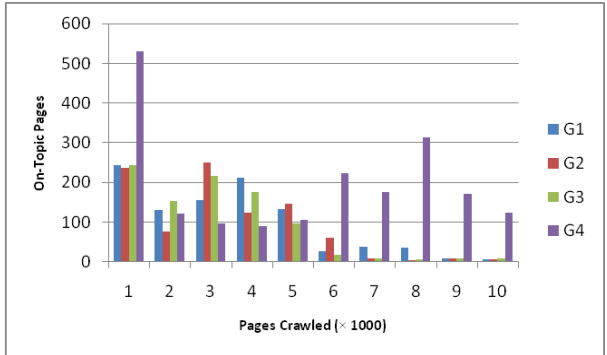
Fig. 5. On-topic retrieved pages (Generic seeds)

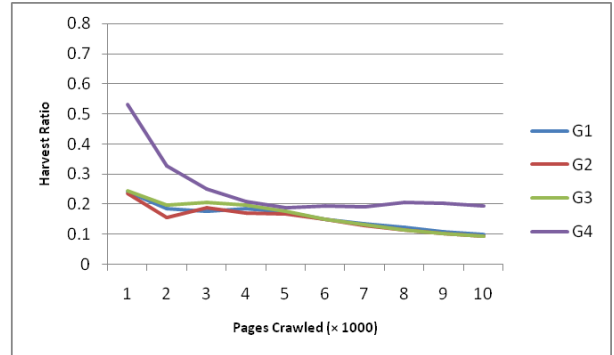
Fig. 9. Harvest Ratio (Generic seeds)

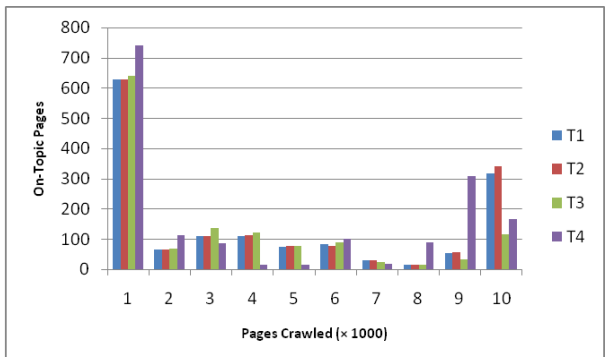
Fig. 6. On-topic retrieved pages (On-topic seeds)

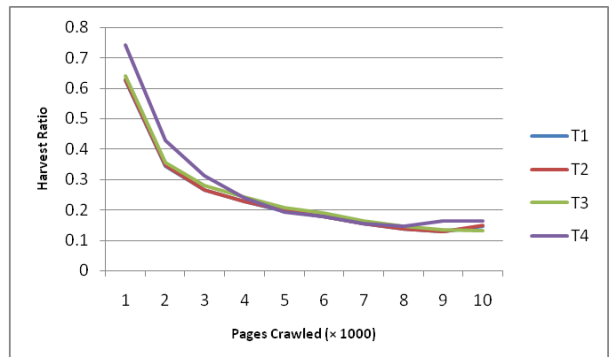
Fig. 10. Harvest Ratio (On-topic seeds)

es, and tries to go through the highly populated regions. Fig. 7 and Fig. 8 show the cumulative number of retrieved on-topic pages. The observable increase in this figure conveys the good performance of the Treasure-Crawler. Different affects of seed types are also apparent in these figures.

As seen in the figures, T4 and G4 show a rather better performance. This shows that increasing the priority score of the fetcher queue items with a lower value will help related URLs to be fetched and downloaded faster, and hence their associated Web pages are primarily analyzed.

Harvest ratio is a significant IR performance metric, and is broadly used in the evaluation of focused Web crawlers. Although the Treasure-Crawler shows a decreasing harvest ratio, its gradual value is still reliable. Considering the effect of the fetcher queue aging factor (as in T4 and G4), the alternate values of other initial parameters, and other system and resource limitations in our research, the harvest ratio of the Treasure-Crawler remains acceptable. (see Fig. 9 and Fig. 10)

## 6. EVALUATION OF TREASURE-CRAWLER SYSTEM

Generally a crawler is expected to retrieve "good" pages. However, to determine what is a good page is not an easy task. Besides, the evaluation of a focused Web crawler can be carried out by considering various criteria, all of which can be categorized as the following:

**Page Importance:** Basically two major types exist in defining an important page; link-based and similarity-based [11]. What we discuss in our focused crawler is similarity-based importance of a page. In this context, similarity is defined as the relevance score of the page to a specific topic in which the crawler specializes. However, measuring and setting thresholds for this score are experimental tasks. Also, the calculation algorithm of this metric is one of the fundamental factors of designing a good crawler.

**Access Speed:** The rate at which a crawler covers a part of the Web defines its speed. This speed depends heavily on the algorithms and data structures it uses. Generally, the average processing time to produce outcomes, whether positive or negative is the important factor to evaluate the speed.

**Repository Freshness:** The Web is rapidly growing and changing. Every Web page's contents change in an unpredictable amount of time interval. To keep up with this change, the crawler has to provide its repository with an algorithm or facility that brings freshness to its contents, meaning that periodically checks the downloaded pages for possible further changes.

**Scoring Function:** To provide an appropriate relevance score for a Web page, the system should employ fast and optimal functions. Using the T-Graph in this regard is a jump towards the optimality of this score. Authors of the Context Graph algorithm, Diligenti et al. [12] have also shown that the average relevance of the downloaded on-topic documents remains consistent and improved, if a sliding window is employed to limit the number of downloads. This concept has not been tested in our system but is proved to improve the performance of such crawler systems that tend to download pages from the Internet.

**Efficiency:** The crawlers require several resources such as network bandwidth to crawl the Web and acquire documents, memory to keep their internal data structures to be used by their algorithms, processor to evaluate and select URLs, and disk space to keep the processed text and fetched links of pages. Having these factors, it should be added that in our prototype implementation, these resources were not considered as an effective factor.

**Recall and Precision:** Recall and precision are the two significant performance measures. There are many other suggested metrics that depend on recall and precision in nature. For example, "the percentage of relevant pages retrieved over time" and "the percentage of papers found as the percent of hyperlinks followed increases" are both estimates of recall. Also "the harvest rate" [13], "the average rank of retrieved pages" [11], and "the average relevance of pages" [12] are all estimates of precision. Srinivasan et al. [14] proposed a performance/cost analysis as a way to gauge the effectiveness of the crawlers against their efficiency. The literature on the performance of focused Web crawlers demonstrates that in all approaches of crawling the Web, as the number of crawled pages increases, the recall and precision values act conversely. Srinivasan et al. [14] also studied several crawling algorithms and have experimentally proven that as the number of crawled pages increases, the average recall of the target pages levels up, while the precision decreases. It must be noted that in a large number of pages, both recall and precision approach a constant level.

In order to present the performance of the Treasure-Crawler, it is compared to the context focused crawler [12]. The context graph algorithm, firstly introduced by Diligent et al., has been the basis for many researches in the domain of Web search crawlers, and many variations have been built based on the concept of context graphs. This approach focuses on the link distance of each Web page from a target document, comprising that if a Web page corresponds (in terms of content similarity of the whole page) to a layer of the graph, the priority of the unvisited link in that page accords to the level number of the graph.

Although the context graph approach is based on the assumption that the documents follow a common hierarchy, it is obvious that the Web pages are not well-organized and homogenously structured. This makes the use of a layered structure inevitable. Fig. 11 compares the performance of the Treasure-Crawler and the context graph algorithm in terms of remaining on topic while crawling the Web. The proposed system signifycantly outperforms the context graph algorithm in terms of the number of retrieved on-topic documents.

In order to obtain the ratio of the relevant to all crawled pages, the recall value (0.50) which is gained from running the system while linked to the DMOZ dataset is applied to the number of retrieved pages. Fig. 12 shows the results, where our system still outperforms the Context Graph algorithm.

In the vast domain of Web crawler technology, there are several common components and algorithms that are frequently used by the designers. Table 7 shows the relevance calculation and classification methods used in the surveyed algorithms in this paper. Although our proposed algorithm hardly uses these methods, they obviously have had a great influence on the crawlers so far. It can be vividly derived from Table 7 that the link-based algorithms are more widely used since downloading the entire content of a Web page puts a huge burden on the search engine server. The Treasure-Crawler employs a novel interpretation of the surrounding text by taking into account the major HTML elements of the page. The classification method, however, highly depends on how a possible approach is adopted. What makes the Treasure-Crawler innovative in this regard is that it totally depends on a custom model in classification and prioritization of pages. Although the performance results are weaker than some other approaches, our proposed algorithm has a truly flexible structure to be enhanced as a future study.

Since all the named systems seem to have close results, it is definitely the matter of choosing the right method. The crawlers are composed of several components, for each of which the designer has many choices.

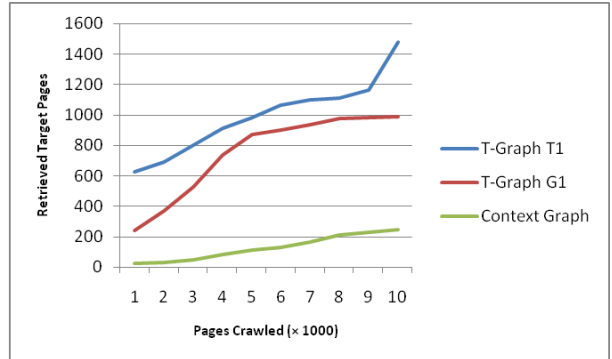

Fig. 11. Comparison of retrieved documents

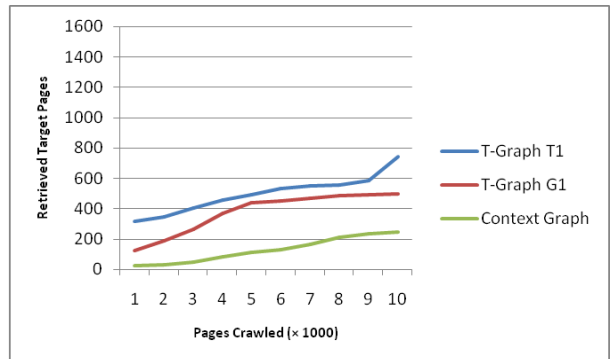

Fig. 12. Comparison of retrieved documents multiplied by recall value (0.5)

TABLE 7

COMPARISON OF DIFFERENT CRAWLERS' RELEVANCE CALCULATION AND CLASSIFICATION METHODS

| Focused Crawler (FC) | Relevance Calculation | | | | Classification | | | | Results | | |
|---|---|---|---|---|---|---|---|---|---|---|---|
| | Whole Content | URL | Anchor Text | Terms Around Links | Custom Method | Naïve Bayes | Decision Tree | Custom Model | Semantic | Recall | Precision | Harvest Ratio |
| CFC | * | | | | * | * | | | | | | 2.5% |
| LSCrawler | | * | * | * | | | * | * | | 60% | | |
| Relevancy CFC | | * | * | * | | | | | | | 22% | |
| Hybrid FC | * | | | | | | | * | | | | 67.2% - 85.5% |
| Meta Search | | | | * | | | | * | | | 50.23% | |
| HAWK | * | * | * | * | | | | * | | | | 60% |
| Intelligent FC | | | | * | * | | | | | | | 80% |
| Modified NB | | * | * | | | * | | * | | | | 33% |
| OntoCrawler | * | | | | | * | * | * | | | 90% | |
| Treasure-Crawler | | * | * | * | * | | * | * | | 50% | 50% | 20% |

Therefore, in order to make a better comparison among this number of systems so far, it is desirable that each fundamental component be compared/evaluated separately. Also, we lack a general evaluation framework with the capability of hosting any Web crawler, which can automatically test and evaluate the different parts of the crawler, namely the relevance calculator and the prioritizer.

In this section, we compared the results from our prototype system with the performance of the context focused crawler. Our system holds the value of 0.5 for both recall and precision which implies that our proposed system outperforms the context focused crawler in terms of these two substantial criteria. Although many improvements have been made on the classic context graph algorithm, no one has tried to remove the strict link distance requirement to make this algorithm more adaptable, before the introduction of T-Graph [8].

### 6.1 Discussion

We deployed a focused Web crawler system based on two main hypotheses. The first was to increase the accuracy in prediction of the subject/topic of an unvisited Web page for which we extracted and utilized specific determinant HTML elements of its parent page. Our results showed the effectiveness of this assumption in the dimensions of a prototype. The second hypothesis was to guide the crawler to find the closest-to-optimal path to harvest the Web with the use of a custom hierarchical data structure called T-Graph. The T-Graph guided the crawler to find highly populated regions of the Web where on-topic pages were found. The experimental results satisfied this assumption as well.

To complete the system evaluation and in order to portray where in the literature of focused crawler subject field our proposed algorithm fits, the Treasure-Crawler was compared to other methods of crawling, namely the Context Focused Crawler [12], LSCrawler [15], Relevancy Context Graphs [16], Hybrid Focused Crawler [17], Meta Search Crawler [18], HAWK [19], Intelligent Focused Crawler [20], Modified Naïve Bayes [21] and the OntoCrawler [22]. Table 7 presented this comparison in terms of the employed algorithm for each module to crawl the Web, as well as the performance results of the mentioned systems. In our method, we attempted to keep the textual comparisons minimal and only relied on simple string functions instead of complicated data mining algorithms. This simplistic way of calculations brought about more realistic and reliable results. We experienced several limitations and restrictions in our implementation, some of which definitely affected the results. To name a few: lack of an ideal parallel or distributed Web server with specific configurations, lack of a high-bandwidth Internet connection with appropriate specifications, lack of a complete list of Dewey entries, and shortcomings of the DMOZ dataset.

It was therefore proposed that any further expansion or implementation of the Treasure-Crawler be carried out by pre-satisfying these constraints. For example, the availability of the complete Dewey system entries will definitely improve the performance of the system in terms of accuracy and preciseness. In our test runs, the DMOZ data set was considered to represent the whole Web. However, the lack of interconnections of Web pages in this data set affected the performance in harvesting its pages from one to another.

Although the main stated assumptions of this research were satisfied, it was obvious that having the above limitations relaxed, we would gain more remarkable results and a highly reliable and flexible innovative method.

### 7. CONCLUSION AND FUTURE DIRECTIONS

Although the page content, hierarchy patterns and anchor texts are satisfactory leads, a focused crawler inevitably needs a multi-level inspection infrastructure to compensate their drawbacks. Unfortunately the current papers overlook the power of such comprehensiveness [23]. Considering these shortcomings, our proposed Treasure-Crawler utilized a significant approach in crawling and indexing Web pages that complied with its predefined topic of interest. The hierarchical structure of the T-Graph guided the crawler to detect and harvest the regions of the Web that embodied a larger population of on-topic pages. The main idea was to detect the topic boundary of an unvisited URL. This was first manually checked to make sure that the HTML parser was correctly fetching the HTML elements of the page. After the necessary textual data in a page was extracted, the system performed one of its major tasks; detecting the topical focus of the unvisited page. The next major task was to assign a score to each URL. This was carried out by use of the pre-constructed but updating T-Graph.

Seed URLs played key roles in the Treasure-Crawler. The generality of the generic seeds as well as the relatedness of on-topic seeds were highly important.

A future direction is in the design of a real search engine: Assuming that a real search engine is to be constructed based on a Treasure-Crawler. There will be the need for a distributed infrastructure which synchronizes several Treasure-Crawlers that specialize in diverse topics. Then an index synchronizer will virtually incorporate these repositories. The search engine then establishes and manages appropriate connections from the user interface to this merged index of the Web.

In comparing and evaluating the Treasure-Crawler

against other focused crawlers, some factors are necessary to be considered. For example, the semantic crawlers should be evaluated in different terms as they are employing a more sophisticated framework – the semantic Web. Another variable that should be taken into account is the number of crawled pages, which should be equal for all the crawlers in the comparison. More importantly, the crawlers in the comparison must focus on a common topic of interest, since the topical communities are absolutely diverse on the Internet. Obviously, if one decides to design a new focused crawler, these diverse factors are of a great help to aim for optimality in focused crawling.

## 7.1 Future R&D Directions

Focused crawling is an exciting research domain, where making any change in any parts of the system will affect the overall performance of the system. This changeability and response is what makes it a challenge for the scientists to continue their research to obtain the desired results. The following changes and additions are just some of many issues that can be taken into account for further developments of the Treasure-Crawler:

• Lemmatization usually refers to doing things properly with the use of a vocabulary and morphological analysis of words, normally aiming to remove inflectional endings only and to return the base or dictionary form of a word, which is known as the lemma. Since there are some inefficiencies or errors in measurements with stemming algorithms, such as over-stemming and under-stemming, it is suggested that a lemmatization algorithm may be used in parallel with the stemming algorithm.

• According to the use of different terminology for similar concepts plus the use of combinatory terms (called conflation) as a typical human behavior, many text materials on the Web might be differently written but convey the same meaning. Regarding this fact, using a thesaurus in order to better detect the topical focus of an unvisited link is suggested. While the system is equipped with such resources, the text similarity measurements will take a semi-semantic way becoming more effective and reliable. For this issue, WordNet [24] is a suggested tool that can be furnished into the system resulting in a more semantic nature.

• As a significant problem in the process of natural languages, word sense disambiguation aims to determine which sense of the word (when it has several meanings) is meant to be used in a sentence. To overcome this issue, several algorithms as well as lexical resources are introduced and can be used. As the text comparison and topic detection are very important processes in a focused Web crawler, arming the system with these capabilities will pragmatically increase the accuracy of the system.

• One of the best known algorithms in prioritizing the links in a Web page is PageRank which takes into account the incoming and outgoing links to and from the Web page in order to calculate the priority score. It is suggested that PageRank (or a modification) may be embedded into the Treasure-Crawler to increase its performance more efficiently.

• A system facility should be present in such a crawler system to control the different behaviors of the repository which is actually a large database. For example, as already included in the current version of this system, every time a Web page is about to be stored, the DB Controller component checks whether the page is already stored or not. If yes, one alternative is that the record is replaced keeping the older version as history. This facility can basically depend on the concept of triggering in order to be implemented.

• If the approach to the construction of T-Graph is chosen to be bottom-up, then the search API of the Google is suggested to be used. In this case, the back crawling approach conducts the search for the parents of each node of the graph. Taking this approach has several benefits as well as drawbacks. Using the Internet is surely slower and requires additional resources. On the other hand, the efficiency of the graph, in terms of quality of the nodes and links, will be increased since the current search engines such as Google use highly intelligent algorithms in classification of pages and topics.

• Although the function to calculate the OSM is defined to be the average of the four similarity values, it is suggested that it assigns a weight to each of those partial similarity values according to their effect. This effect weight should be calculated experimentally to gain the minimum error rate. For example, we can compare parts of all the nodes in a level and then normalize the values by dividing each by the sum of all (ratio). This will be an optimization problem called "constraint optimization", which means setting these weights (of SIMs) to reach the minimum error where the summation of all values will be 1, and each of them is between 0 and 1.

• It is considered that the priority score of an unvisited link is the inverse of the least number of links that should be traversed to reach the target level in the T-Graph. Another suggested solution to calculate the priority in a more effective way is to consider the level number of all the similar nodes as well as their link distance to reach the target level simultaneously. Also we can add the OSM of each node into the formula for a better and more logical outcome.

　　The above given directions are only some of many ideas that can improve further versions of the Treasure-Crawler. However, it will be very hard work to compete with other sophisticated search engines. According to the unstoppable growth of the Web, focused crawlers are currently forming an interesting research area.

Finally, the theory of using significant HTML elements of a parent Web page in predicting the subject of its child pages is unlimitedly flexible. There are diverse approaches to classify Web pages into specific categories of human knowledge; however a researcher may prefer to design custom methods/models as proposed in this paper. The use of a custom designed structure of exemplary documents is one other subject to investigate further. This structure is what differentiates focused crawlers from topical crawlers, and what makes focused crawlers more widely researched on.

## ACKNOWLEDGEMENT

This R&D work has been conducted at the National University of Malaysia as the author's master's thesis.

## REFERENCES

[1] A. Seyfi "A Focused Crawler Combinatory Link and Content Model Based on T-Graph Principles," 2013

[2] M. d. Kunder. (2011, 3 January 2011). *The size of the World Wide Web*. Available: http://www.worldwidewebsize.com

[3] Frank. (2011, 20 May 2011). *Let's Do Dewey*. Available: http://frank.mtsu.edu/~vvesper/dewey2.htm

[4] OCLC. (2011, 20 August 2010). *Dewey Services at a glance*. Available: http://www.oclc.org/dewey/

[5] M. F. Porter, "An Algorithm for Suffix Stripping," *Program: electronic library and information systems,* vol. 14, pp. 130-137, 1980.

[6] H. Kargupta*, et al.*, *Next Generation of Data Mining*, 1 ed.: Chapman & Hall/CRC, 2008.

[7] A. Passerini*, et al.*, "Evaluation Methods for Focused Crawling," in *AI*IA 2001: Advances in Artificial Intelligence*. vol. 2175, F. Esposito, Ed., ed: Springer Berlin / Heidelberg, 2001, pp. 33-39.

[8] A. Patel, "An Adaptive Updating Topic Specific Web Search System Using T-Graph," *Journal of Computer Science,* vol. 6, pp. 450-456, 2010.

[9] DMOZ. (2011, 10 March 2011). *Open Directory Project*. Available: www.dmoz.org

[10] Yahoo! (2011, 10 April 2011). *SiteExplorer*. Available: https://siteexplorer.search.yahoo.com

[11] F. Menczer*, et al.*, "Evaluating Topic-Driven Web Crawlers," in *SIGIR'01*, New Orleans, Louisiana, USA, 2001.

[12] M. Diligenti*, et al.*, "Focused Crawling Using Context Graphs," in *26th International Conference on Very Large Databases, VLDB*, Cairo, Egypt, 2000, pp. 527-534.

[13] S. Chakrabarti*, et al.*, "Focused Crawling: A New Approach to Topic-Specific Web," in *The 8th World-wide web conference (WWW8)*, 1999.

[14] P. Srinivasan*, et al.*, "A General Evaluation Framework for Topical Crawlers," *Information Retrieval,* vol. 8, pp. 417-447, 2005.

[15] M. Yuvarani*, et al.*, "LSCrawler: A Framework for an Enhanced Focused Web Crawler Based on Link Semantics," in *IEEE/WIC/ACM International Conference on Web Intelligence*, 2006.

[16] C.-C. Hsu and F. Wu, "Topic-specific crawling on the Web with the measurements of the relevancy context graph," *Information Systems,* vol. 31, pp. 232-246, 2006.

[17] M. Jamali*, et al.*, "A Method for Focused Crawling Using Combination of Link Structure and Content Similarity," in *IEEE/WIC/ACM International Conference on Web Intelligence*, 2006.

[18] J. Qin*, et al.*, "Building Domain-Specific Web Collections for Scientific Digital Libraries: A Meta-Search Enhanced Focused Crawling Method," in *Joint ACM/IEEE Conference on Digital Libraries*, Tucson, Arizona, USA, 2004.

[19] C. Xiaoyun and Z. Xin, "HAWK: A Focused Crawler with Content and Link Analysis," in *e-Business Engineering, 2008. ICEBE '08. IEEE International Conference on*, 2008, pp. 677-680.

[20] D. Taylan*, et al.*, "Intelligent Focused Crawler: Learning which links to crawl," in *Innovations in Intelligent Systems and Applications (INISTA), 2011 International Symposium on*, 2011, pp. 504-508.

[21] W. Wang*, et al.*, "A Focused Crawler Based on Naive Bayes Classifier," in *Intelligent Information Technology and Security Informatics (IITSI), 2010 Third International Symposium on*, 2010, pp. 517-521.

[22] O. Jalilian and H. Khotanlou, "A New Fuzzy-Based Method to Weigh the Related Concepts in Semantic Focused Web Crawlers," in *Computer Research and Development (ICCRD), 2011 3rd International Conference on*, 2011, pp. 23-27.

[23] R. Chen, "An Enhanced Web Robot For The Cindi System," 2008.

[24] U. Princeton. (2011, 6 April 2011). *WordNet*. Available: http://wordnet.princeton.edu/wordnet/